\documentclass[10pt]{article}
\usepackage{epsfig,graphicx}
\usepackage{times,ch2005}

\def\beq{\begin{equation}}
\def\eeq#1{\label{#1}\end{equation}}
\def\eeqn{\end{equation}}

\def\beqa{\begin{eqnarray}}
\def\eeqa#1{\label{#1}\end{eqnarray}}
\def\eeqan{\end{eqnarray}}

\def\Dslash{\not{\hbox{\kern-4pt $D$}}}
\def\dslash{\not{\hbox{\kern-2pt $\del$}}}

\newcommand{\tev}{\ensuremath{\mathrm{\,Te\kern -0.1em V}}\xspace}
\newcommand{\gev}{\ensuremath{\mathrm{\,Ge\kern -0.1em V}}\xspace}
\newcommand{\mev}{\ensuremath{\mathrm{\,Me\kern -0.1em V}}\xspace}
\newcommand{\kev}{\ensuremath{\mathrm{\,ke\kern -0.1em V}}\xspace}
\newcommand{\ev}{\ensuremath{\mathrm{\,e\kern -0.1em V}}\xspace}
\newcommand{\gevc}{\ensuremath{{\mathrm{\,Ge\kern -0.1em V\!/}c}}\xspace}
\newcommand{\mevc}{\ensuremath{{\mathrm{\,Me\kern -0.1em V\!/}c}}\xspace}
\newcommand{\gevcc}{\ensuremath{{\mathrm{\,Ge\kern -0.1em V\!/}c^2}}\xspace}
\newcommand{\mevcc}{\ensuremath{{\mathrm{\,Me\kern -0.1em V\!/}c^2}}\xspace}

\def\mus  {\ensuremath{\rm \,\mus}\xspace}

\def\mus        {\ensuremath{\,\mu{\rm s}}\xspace}    

\begin{document}


\Title{Review of the Solar Array Telescopes}
\bigskip


%
\label{SmithStart}

%
\author{ David A. Smith\index{Smith, D.A.} }

%
\address{Centre d'Etudes Nucl\'eaires de Bordeaux-Gradignan - CNRS/IN2P3 \\
26, All\'ee du Haut Vigneau\\
F-33175 Gradignan, France \\
}

\makeauthor
\abstracts{
For several years the only experiments sensitive to astrophysical gamma rays with energies beyond
the reach of EGRET but below that of the Cherenkov imaging telescopes have been the "solar tower" detectors.
They use $>2000$ m$^2$ mirror areas to sample the Cherenkov wavefront generated by $<100$ GeV gamma rays, obtaining
Crab sensitivities of more than 6$\sigma$ in one ON-source hour. I will review the history of the solar tower
Cherenkov experiments from 1992 to the present and their key design features. I will describe some successful
analysis strategies, then summarize the principal results obtained.
\\
\em{Contribution to the proceedings of "Towards a Network of Atmospheric Cherenkov Detectors VII", 
 27-29 April 2005 - Ecole Polytechnique, Palaiseau, France.}  
 http://polywww.in2p3.fr/actualites/congres/cherenkov2005/
}

\section{The Quest to Open the 50 GeV Gamma Ray Window}

In 1992 we knew that the window between EGRET and Whipple is particularly rich. Was there a way to glimpse
into that window without waiting 10+ years for the next space mission (GLAST), or 10+ years to move 
``Towards a Large Atmospheric Cherenkov Detector''? At that time stereo (HEGRA) and fine-pixel cameras (CAT) were
only just being designed and built, although big mirrors (Whipple) were already well-established.

\begin{figure}[htb]
\begin{center}
\epsfig{file=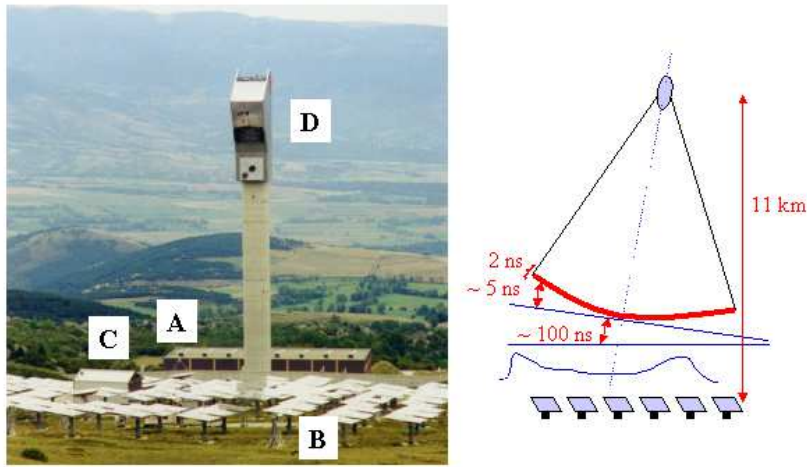,height=10cm}
\vspace*{-2.5cm}
\caption{Left: The Th\'emis solar facility. Beneath A is one of the 7 ASGAT telescopes.
B is one of the 18 Themistocle telescopes.
C is the hangar housing the CAT imager. D is where the CELESTE secondary optics and
counting house were.  Right: For a source
20 degrees from the zenith and a heliostat 100 meters from the array center,
the wavefront inclination retards the Cherenkov light by $\sim 100$ ns, and the
wavefront sagitta is $\sim 5$ ns. The wavefront thickness (pulse duration) is
$\sim 2$ ns. The curve is an artist's impression of the Cherenkov density across
the light pool, described as an `Arizona mesa' in the text.}
\label{fig:Smith-fig1}
\end{center}
\end{figure}

Wavefront sampling was a proven alternative to imaging: 
the THEMISTOCLE \cite{themistocle} and ASGAT \cite{asgat} experiments at Th\'emis, in
the French Pyrenees, had both detected the Crab (visible in figure \ref{fig:Smith-fig1}). The PACT array in Madhya Pradesh, India,
continues on that path \cite{india}. Imaging performance was better, but we had several reasons to believe that at
lower energy, wavefront sampling could have advantages \cite{SanMiniato}.

After the 1973 oil crisis, solar tower facilities were built in several countries to
generate electricity, or as high temperature, high heat research facilities. Their gigantic mirror areas, on
hundreds, even thousands, of alt-azimuth mounts spread over areas as big or bigger than a Cherenkov light
pool made every ground-based gamma ray astronomer dream \cite{Castagnoli, Danaher}. 
A key breakthrough was when Tumay T\"umer proposed
to place secondary optics and a photomultiplier camera near the top of the tower at Solar-I, near Barstow,
California \cite{Tumay}. 
Tumay persuaded many of us to give it a go (see e.g. \cite{earlyStacee,pareEarly}).

Four Solar Tower experiments have been built: Most of the original SOLAR-I group moved to Sandia Laboratory in
Albuquerque, New Mexico, and runs the STACEE experiment \cite{staceeNim}. The CELESTE experiment, shown in
figure \ref{fig:Smith-fig1}, ran at Th\'emis \cite{pareNim}. 
The GRAAL experiment (63 heliostats for 2500 m$^2$, at Almeria, Spain) forewent
secondary optics at the price of a $250 \pm 110$ GeV energy threshold, 
yet succeeded at detecting the Crab nebula \cite{graal}. Their discussion of the effects
of a small field-of-view is rich.
The CACTUS experiment, at what is now called SOLAR-II, currently has twice
the mirror area of STACEE and CELESTE and aims to expand (P. Marleau \& M. Tripathi, these proceedings). 
This article reviews how we exploited our ten year window of opportunity, concentrating mainly on the STACEE
and CELESTE experiments.

\vspace*{-0.9cm}


\section{Wavefront Sampling with Solar Mirror Arrays}
\subsection{The ABC's}

Figure \ref{fig:Smith-fig1} illustrates the principle of a solar tower experiment: 
the light intensity and arrival time are
measured at dozens of points across the lightpool. As for most experiments, the two main challenges are
triggering, and analysis.
Equation \ref{Eth} for the minimum threshold energy is a basic paradigm for all atmospheric cherenkov
detectors \cite{weekes}, 
\begin{equation}
E_{th} = \sqrt{\Omega\phi\tau \over A\epsilon}.  
\label{Eth}  
\end{equation}

STACEE and CELESTE have primary mirror areas $A$ over 2500 $m^2$, over a hundred times more than CAT's 16
$m^2$.  Since CAT's energy threshold is $E_{th}\sim 250$ GeV, one might expect STACEE \& CELESTE to go below
25 GeV.  But CAT pushed the electronic integration time $\tau$, and the pixel size $\Omega$, 
as far as they could go to obtain roughly the same performance as Whipple with its 75 $m^2$ mirror. 
$\tau$ is driven by the $\sim 3$ ns width of the Cherenkov pulse for a photon shower. 
For a solar tower experiment $\Omega$ is also the telescope field-of-view. STACEE \& CELESTE
have $\Omega = \pm 5$ mrad, driven by the apparent size of a gamma shower.
The photon conversion efficiency $\epsilon$ is
dominated by the bialkili photocathodes generally in use -- the solar tower optics further halve $\epsilon$
as compared to imagers. The intensity of the night sky background light, $\phi$, 
varies by almost a factor of two between sites. The Cherenkov signal is roughly 4 photoelectrons per heliostat
for a 50 GeV gamma ray in the case of CELESTE.

In STACEE and CELESTE the design principle was that one heliostat per channel, with one FADC per channel, is
the best way to optimize the Cherenkov to night sky ratio. CACTUS takes a different twist: they say that the
power of Flash ADC's can be better brought to bear. In particular, they increased the number of
heliostats without a proportional increase in electronics costs by having larger $\Omega$. 
Each phototube (like for GRAAL) sees many heliostats, but also integrates more night sky
background (`nsb' $= \phi A \Omega \epsilon$ is of the order of $10^9$ photoelectrons per second). 
The Flash ADCs allow them to identify the signals from individual heliostats, and to
integrate the nsb only over the duration of that signal. The handicap of increased nsb is diminished by their
having a large $\sqrt{N_{heliostats}}$. Field-of-view is discussed further, below.

For all the solar tower experiments, the earth's rotation makes the optical path lengths change
constantly (about 2 ns per minute). Considerable effort was invested in the programmable delays
needed to keep $\tau$ small. High rates (kHz for CELESTE, MHz for STACEE, due to different
trigger concepts) before imposing the trigger coincidences require
the digital delays to have low deadtimes. Long delays (hundreds of ns) make signal deformation an
issue for analog delays. The software to drive the delays and to identify the FADC readout windows containing
the Cherenkov pulse needs to be precise. 

Flash ADC's (or their close cousins) have become standard for modern Cherenkov experiments, but
the solar tower experiments were the first to use them, requiring a substantial development effort.
They have proven to be extremely useful.

\subsection{A brief chronology}
{\bf 1994-1995:} Cherenkov flashes on four heliostats were recorded at Solar-I, 
to begin characterising the heliostat optics
and tracking. Secondary optics is a plastic Fresnel lens \cite{earlyStacee}. 
CELESTE design work essentially finished \cite{pareEarly}.
\newline
{\bf 1995-1996:} STACEE moves to Sandia, makes further Cherenkov measurements \cite{staceeSandia}.
CELESTE proposal written \cite{proposal}, and the industrial heat receiver is removed from the top
of the Th\'emis tower. 
\newline
{\bf 1996-1997:} Six Th\'emis heliostats record Cherenkov flashes, using two anti-aircraft mirrors
for secondary optics \cite{berrieNim}. 
\newline
{\bf 1997-1998:} CELESTE optics in place. With temporary electronics, a $5.6\sigma$ Crab excess
near 100 GeV is obtained in 3.5 hours of data \cite{texas}. Eric Par\'e, a founder of the field, is
killed in a car crash. 
\newline
{\bf 1998-2000:} GRAAL Crab detection, using 64 heliostats, no secondary mirrors, and simple
electronics \cite{graal}. STACEE Crab detection \cite{staceeCrab}, 
with a 32-heliostat array \cite{staceeNim}. CELESTE 40 heliostat array completed \cite{pareNim},
Crab and Mrk421 detected \cite{celesteCrab}.
\newline
{\bf 1991-2002:} STACEE measures Mrk 421 flux with a 48 heliostat array \cite{stacee421}, pushes
forward to final 64 heliostat configuration, and begins searching for other blazars \cite{staceeWcom}.
\newline
{\bf 2002-2004:} Solar-II reborn as CACTUS. Conjunction of Saturn with the Crab stops standard candle
studies for one season. CELESTE extends to 53 heliostats but Th\'emis weather keeps getting worse.
A new analysis achieves $5.8\sigma/\sqrt h$ Crab sensitivity \cite{philippe,hakima}. 
Re-analysis of old data gives a signal on Mrk 501 (E. Brion, these proceedings). CELESTE dismantles
in June 2004 while STACEE continues observations and improvements.


\section{Gamma Reconstruction and Hadron Rejection}
\subsection{What we thought we'd be doing}

As mentioned, wavefront sampling was pioneered by the 7 large ASGAT mirrors (\cite{asgat}) and the 18 small
THEMISTOCLE mirrors (\cite{themistocle}). The THEMISTOCLE energy threshold was a few TeV, at which energies the
electromagnetic cascade is a line-like Cherenkov source high in the sky, making the
wavefront essentially conical. Sampling the arrival times allowed a fit to a cone, with the
cone axis yielding the primary particle direction. Hadrons fit the conical hypothesis poorly, providing
rejection.

ASGAT had smaller $\Omega$ (20 vs 40 mrad) and larger $A$ (40 m$^2$ vs 0.5 m$^2$) leading to an energy reach
down to a few hundred GeV. At these lower energies, showers are shorter making wavefronts rounder. For ASGAT
the showers were still somewhat conical, but the effectiveness of the conical fit was diluted, compromising
ASGAT's sensitivity.

The solar tower detectors see showers with several 10's of GeV -- the shower is a blob and the wavefront spherical.
Fitting the arrival times gives the center of the sphere.
The light distribution on the ground ressembles a flattish mountain, with a
``cliff'' 120 meters from the center (imagine an Arizona mesa). 
The ``Cherenkov ring'' of the air shower makes a ridge at the edge of the cliff.

Early attempts to obtain a gamma ray direction used the pulseheight distribution on the ground to search for the
gamma ray's (extrapolated) impact point -- Monte Carlo studies gave hope that one could use the edge, and perhaps the ring, 
to find the center of the light pool. The impact point on the ground and the shower position in the sky together yield the
gamma ray direction, and a direction cut would discriminate between the point source and the isotropic
cosmic ray background. 

The small field-of-view messes this up: for showers impacting away from the center of the heliostat field,
the Cherenkov light is near the heliostat optic axis near the impact point, and off axis for heliostats far away. 
The observed pulseheights per heliostat are badly distorted (the `Arizona mesa' is eroded), 
and the direction resolution is comparable to the field-of-view, that is, not useful.
This same effect also makes event-by-event energy determination difficult -- the total pulseheight depends not just
on the primary energy but on the impact parameter. Energy reconstruction was accomplished nevertheless
(see figure \ref{fig:Smith-fig2}and \cite{fredSpectra,staceeSpectra}), as discussed below.

Another simple idea is to use the goodness-of-fit (e.g. a $\chi^2$ cut) to distinguish between proton and gamma
showers. That works poorly at low energies because the pulses on the individual heliostats are so close to the
fluctuations of the night sky background that the proton vs gamma differences get smeared out. 

STACEE and CELESTE came to recognize these difficulties and to find ways around them over the years. CELESTE's best
variable in the early days was $\sigma_{group}$ -- five groups of heliostats were summed together, and the dispersion
of the five charge sums provided gamma/proton separation \cite{celesteCrab}. STACEE achieved initial success with the
classic approach outlined above \cite{staceeCrab}.

\subsection{Field-of-view, and Analog Sums} 
The choice of $\Omega$ in a solar tower experiment has many consequences.
Even the energy threshold, $E_{th}$ (equation \ref{Eth}), is not simple: STACEE \& CELESTE
point not at the source (at infinity, called ``parallel heliostat pointing'') but at shower maximum,
about 11 km above the site (``convergent'' or ``canted'' pointing, see for example \cite{herault}). 
This increases the Cherenkov light
collection, decreasing $E_{th}$, but also decreases the effective area of the telescope: imagers have
areas of a few $10^5$ m$^2$ compared to our few $10^4$ m$^2$. I speculated, above, that CACTUS, with
large $N_{heliostats}$ and $\Omega$ might win back some m$^2$ without losing GeV.

According to the so-called ``small field-of-view problem'' (small $\Omega$), the tails of 
proton showers recorded by CELESTE \& STACEE get cut, making the background events look like
signal events in some variables: $\sigma_{group}$ would work better if heliostats were seeing
more of the tails and the sub-showers.
CELESTE dedicated 12 of the 53 heliostats to a ``veto'' role -- by pointing away from the nominal shower core,
they effectively extend the field-of-view so that proton showers can be tagged. This doubled
sensitivity \cite{hakima} . STACEE applies similar tactics \cite{staceeWcom}.

A $\sigma_{group}$-like approach should work better for CACTUS, and direction \& energy reconstruction could
also be easier. The canting altitude could be higher, thereby increasing the gamma ray effective area.
But in the end, the ``problem'' isn't that bad, since analog sums also provide good cosmic ray rejection.

As mentioned, even if small Cherenkov pulses are gleaned from the night sky light, 
their arrival times and amplitudes are smeared. 
A coherent sum greatly improves the signal-to-noise ratio: CELESTE obtains the lowest energy threshold of the
different experiments using {\em analog} trigger sums. 
Summing requires an assumption as to the shower core location. In
analysis, CELESTE turned that around by looping over a grid of possible positions, and retaining the one giving the
most gamma-like sum, that is, the largest, narrowest peak possible \cite{philippe,hakima}. 

Energy reconstruction, and thus measurement of differential spectra, 
has been achieved by both STACEE \& CELESTE, as shown in figure \ref{fig:Smith-fig2}.
In \cite{fredSpectra} the spherical fit was used to estimate the shower position. 
Monte Carlo simulations provided
tables of observed charge versus energy and shower position, from which functions providing
energy based on charge and position were built. The energy resolution obtained is about 30\%, with a
bias of less than 5\%. 
The STACEE points in the figure are from early 2004 \cite{staceeSpectra}, when Mrk 421
was in `medium' and `high' states, and 
represent a significant breakthrough in STACEE's data analysis.
(The STACEE `butterfly' (or `bowtie') from \cite{stacee421} is
not a true spectral measurement: it represents a range of assumed power laws used to
place their integral measurement on a differential plot. Mrk 421 was in a very high
state for these 2001 data.)

\subsection{Systematic biases}
As stated, single PMTs see 1 GHz of night sky photoelectrons. After grouping together and applying
few photoelectron thresholds, high rates are sent to programmable delays with finite deadtimes. 
Small differences in the nsb between ON and OFF lead to trigger biases, and also to biases
in the Flash ADC data. Care is taken to remove the biases (via offline pulseheight cuts,
and ``software padding'' \cite{celesteCrab} 
and ``field brightness corrections'' \cite{staceeWcom}), at some price in energy threshold.

Cosmic rays dominate the trigger rates, which are typically 25 Hz for CELESTE and 7 Hz for STACEE.
The STACEE Atmospheric Monitor (``SAM'') is an
optical telescope that tracks the same source and provides nsb and stellar extinction information.
CELESTE used the bright star in the field-of-view of Mrk 421 to similar ends.
CELESTE also built a
LIDAR on the site, though its usefulness was ultimately limited by its complexity of operation
and interpretation. 
CELESTE was bothered by a more-or-less seasonal variation in that rate, which tended to
slow the cosmic ray rate to $<15$ Hz, mostly in the springtime. The rate changes are correlated with increased
attenuation of starlight, attributed to aerosols.

Sub-nanosecond timing-in of the signals from the many heliostats is critical to success.
STACEE built a laser flashing system for this purpose \cite{staceeLaser}. CELESTE used
the system installed by the THEMISTOCLE group \cite{themistocle}. 

\section{Results}
CELESTE had better sensitivity but STACEE has more clear nights. In terms of overall results
(detections, and upper limits) the two groups are similar.
\subsection{Crab Nebula Flux and Pulsar Search}
All four experiments have detected the Crab nebula. 
\newline
{\bf GRAAL} obtained a $4.5\sigma$ excess from the Crab in 7.2 hours, at $250 \pm 110$ GeV,
for a sensitivity of $2\sigma/\sqrt h$. Their flux point is $\sim 2\sigma$ above the `consensus'
curve of measurements and models, perhaps due to an underestimation of their acceptance \cite{graal}.
\newline
{\bf STACEE} reported $6.75\sigma$ in 43 ON-source hours (=$1\sigma/\sqrt h$) 
with the 32-heliostat array in \cite{staceeCrab}, with an energy threshold of $190 \pm 60$ GeV.
They reported $2.5\sigma/\sqrt h$ in \cite{kildea}.
They quote an energy threshold of 140 GeV, with sensitivity over the range 50 to 500 GeV \cite{stacee421}.
\newline
{\bf CELESTE} had $2.1\sigma/\sqrt h$ with an early analysis of 40-heliostat data,
which increased to over $3\sigma/\sqrt h$ using the double-pointing configuration \cite{celesteCrab}.
Since then, analysis and detector improvements have increased that to between $5.8\sigma/\sqrt h$ 
\cite{philippe} and $6.5\sigma/\sqrt h$ \cite{lavalle}, depending on the datasets and some
analysis choices. The new analysis applied to the old data gives 4 and $5\sigma/\sqrt h$ for
single and double pointing (E. Brion, these proceedings). 
CELESTE quoted an energy threshold of $60 \pm 15$ GeV in \cite{celesteCrab}. Stellar photometry
studies of optical throughput and atmospheric extinction led us to degrade that, whereas the
`analog sum' analysis described above wins some of it back. Our thresholds straddle 100 GeV, again
depending on the specifics of a given analysis.
\newline
{\bf CACTUS} has a separate contribution to these proceedings (Marleau and Tripathi).

To put the above numbers in perspective: if you're not already familiar with the widespread
``sensitivity of gamma ray telescopes'' plot, you can see {\em e.g.} figure 1 of
http://doc.in2p3.fr/themis/CELESTE/PUB/SmithDolomieu.ps.gz, where the ``CELESTE/STACEE''
curve is centered near 100 GeV, at about $0.1$ Crab. That curve came from \cite{proposal}.
For 50 hours (as in the plot) and $6\sigma/\sqrt h$ one would obtain a $5\sigma$ signal
for a source ${5 \over 6\sqrt 50} = 0.12 $ Crab, which means that we did pretty much what we said
we would back in 1995. 

{\bf Crab Pulsar:} 
High energy pulsar studies bring insight into a range of problems, 
from the fundamentals of the pulsars themselves
to a better mastery of stellar evolution in the Galaxy and the cosmic ray energy budget. 
EGRET gave us a glimpse: several pulsars were seen, 
at least two of which showed no clear high energy cut-off (Crab, and PSR B1951+32) \cite{fierroThesis}. 
CELESTE, with sensitivity down to 30 GeV, had high hopes to detect the Crab pulsar, 
in order to better constrain the cut-off energy. 
Some data sets had intriguing phasograms \cite{manu}, but in the end the limit in
\cite{celesteCrab} is the most reliable result: if the EGRET power law is taken to have an exponential cutoff, 
$e^{-E/E_0}$, CELESTE obtained $E_0 < 26$ GeV. STACEE provided an upper limit at higher energy in 
\cite{staceeCrab}, using somewhat different assumptions. 
CELESTE also obtained a limit for a pulsed signal from
PSR B1951+32 \cite{manu}.

\subsection{Blazar Detections and Upper Limits}
\begin{figure}[htb]
\begin{center}
\epsfig{file=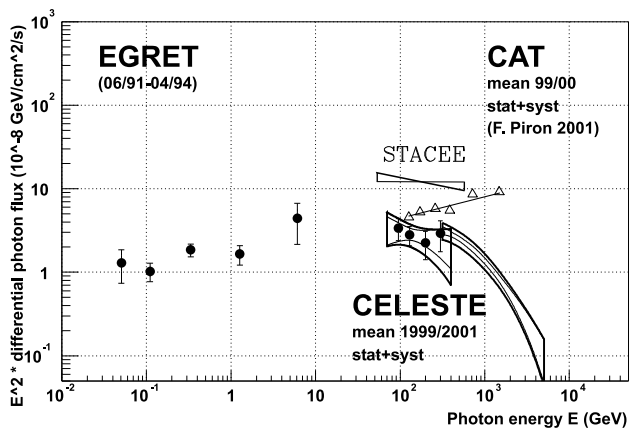,height=14cm}
\vspace*{-7.8cm}
\caption{Spectral measurements of Mrk421 straddling 100 GeV, at three different epochs. 
CELESTE (solid dots, \cite{fredSpectra}) and
STACEE (open triangles, \cite{staceeSpectra}) both reconstruct energy on an event-by-event basis,
providing a differential flux. The STACEE butterfly was an integral measurement for which a
range of power laws were assumed \cite{stacee421}. }
\label{fig:Smith-fig2}
\end{center}
\end{figure}

{\bf Mrk 421} is often as bright as the Crab. Spectral results from STACEE \cite{stacee421} and
CELESTE \cite{fredSpectra} (see figure \ref{fig:Smith-fig2}) were discussed above. Light curves
are in \cite{stacee421,kildea} and \cite{celeste501}.
\newline
{\bf Mrk 501} was detected by CELESTE (E. Brion, these proceedings). 
Data was recorded during six different new moon periods, 
with an overall excess of $2.9 \sigma$.
May and June 2000 stand out, with $4.9\sigma$ total. 
During that period ASM on RXTE shows a slight increase in 2-12 keV X-ray activity above a very faint baseline -- that is, there is 
essentially no detected Xray activity. CELESTE chose to interpret the excess as a mini-flare with an intensity
at the $1 \over 5$ Crab level.
\newline
{\bf 1ES 1426+428} gave STACEE a preliminary $2.9\sigma$ excess in 2003, and $1.1\sigma$ in 2004 \cite{kildea}.
CELESTE reports an upper limit at the $1 \over 5$ Crab level in \cite{celeste501,hakima}. A weak signal at
our energies is favored by Whipple \cite{krennerich} and HESS \cite{hessIR} results indicating 
that the extragalactic infrared density is lower than previously believed.
\newline
{\bf 1ES 0219+42 = 3C66A} gave STACEE a preliminary $2.8\sigma$ excess for 16.9 hours of data \cite{kildea}.
CELESTE derived an upper limit in \cite{rlg} (but only 2.3 hours of data and a preliminary analysis).
{\bf 1ES 2344+514} was searched for in the same thesis. 
\newline
{bf W comae} gave a particulary interesting upper limit for STACEE \cite{staceeWcom}, that 
excludes some proton models, and models invoking inverse Compton scattering of off external
(i.e. not synchrotron) target photons, although the constraint is somewhat weakened by the non-simultaneity
of the STACEE, Beppo SAX, and EGRET data sets.
\newline
{\bf 3C 454.3} and {\bf 3EG 1835+35} were both searched for by GRAAL \cite{graal}.
The former is a distant quasar (z=0.859), amongst the brighter seen by EGRET. The latter
is the bright high galactic latitude EGRET source, that has been associated with a neutron
star since GRAAL ended \cite{3EG1835}.

\subsection{SUSY searches, Supernovae, Gamma Ray Bursts}
In \cite{lavalle}, CELESTE obtained an upper limit from 6.5 hours of observation of the Andromeda galaxy, M31.
M31 has a large mass-to-light ratio and is close to earth. 
The non-detection is interpreted in terms of SUSY neutralino
annihilation, excluding some non-standard scenarios. 
Draco is another interesting Dark Matter candidate, to which CACTUS
has been devoting efforts.

Supernovae remnants were among the original goals of the Solar Tower experiments: EGRET 
extrapolations and some models made detection seem plausible for a few (e.g. IC 443). 
However, as their faintness became apparent (with Cas A a case in point \cite{CasA}), they
fell by the wayside.

STACEE reacts to GRB alerts, and has tracked seven so far. The last thesis underway in
CELESTE explores the possibility of measuring  100 GeV diffuse galactic gamma rays
(R. Britto, these proceedings).

\section{Conclusions}
In 1992 we thought that opening the 100 GeV energy range, between EGRET and Whipple/CAT/Hegra, 
would give fast access to the hundreds of sources seen by EGRET. 
Now, over a decade later, HESS has shown us that the key is {\em not} just to shift in energy range but
especially to improve in sensitivity. The many sources we hoped to see are indeed there, 
but at the $< {1 \over 20}$
Crab level rather than $> {1 \over 5}$. 
In 1992 we also thought that wavefront sampling might have advantages to complement imagers at low energies.
HESS has since confirmed that imaging is {\em very} powerful. 

The solar tower experiments have made the first flux meaurements of the three canonical Cherenkov sources
in this new energy window (Crab, Mrk 421, Mrk501). Upper limits (weak detections?) for a few interesting
blazars, amongst which are W Com and 1ES 1426+428, have been obtained. 
M31 provided a rare constraint on astrophysical SUSY. 
Our Crab pulsar limit is likely to remain the best until GLAST is launched.

The solar tower experiments were successful in spite of a range of adverse conditions, ranging from
the political (STACEE's efforts to convince Commonwealth Edison to let them stay at Solar-II were
nearly as epic as VERITAS's fight to stay on Mt. Hopkins), to the environmental (the number of cloudy
nights at Th\'emis, but also at Sandia, has been getting consistently worse), and to the astrological
(the conjunction of Saturn with the Crab was annoying).

A success not to be overlooked is the training acquired by our students and postdocs,
some of whom have gone on to key roles in HESS and VERITAS, or other fields of physics.
Pioneering the solar tower approach required deep thought about the atmospheric Cherenkov
technique, and the ability to tackle tough experimental problems. Roughly 20 PhD theses
have been written.

What future for wavefront sampling, especially with the new imager arrays making such a splash?
The 2000 heliostats at Solar-II still make some of us dream, and the (small!) CACTUS team is 
implementing some new ideas, but imager arrays will surely be the linchpin of future detectors. 
Could imager sensitivity be enhanced by adding ASGAT-like telescopes?  
The limiting factors are i) primary electrons (or cosmic rays fragmenting almost completely to a $\pi^o$) 
and ii) angular resolution.
Timing arrays give a good measure of the shower core position, which is on average 1 radiation length
higher in the sky for electrons than for gamma rays. 
There could be ideas to explore here.

\section*{Acknowledgments}
Claude Ghesqui\`ere passed away during the preparation of this article (October 2005).
He was one of the key founders of Th\'emistocle, and thus one of the very first high energy
particle physicists to switch over to the Th\'emis solar array and to atmospheric Cherenkov 
light as a means to do gamma ray astronomy. He and I built a nifty muon detector for CAT together...
This article is dedicated to his memory.

%
\label{SmithEnd}
 
\end{document}